\documentclass[12pt]{article}

\begin{document}

\thispagestyle{empty}
\renewcommand{\thefootnote}{\fnsymbol{footnote}}


\*\vspace{1.8cm}

\begin{center}
{\bf\Large Renormalization Group Approach to the Beam-Beam 
Interaction in Circular Colliders} 

\vspace{1cm}

Stephan I. TZENOV\\
{\it Plasma Physics Laboratory, Princeton University,
Princeton, New Jersey  08543}\\
\end{center}

\vfill

\begin{center}
{\bf\large   
Abstract }
\end{center}

\begin{quote}
Building on the Renormalization Group (RG) method the beam-beam 
interaction in circular colliders is studied. A regularized symplectic 
RG beam-beam map, that describes successfully the long-time asymptotic 
behavior of the original system has been obtained. The integral of 
motion possessed by the regularized RG map has been used to construct 
the invariant phase space density (stationary distribution function), 
and a coupled set of nonlinear integral equations for the distributions 
of the two colliding beams has been derived.
\end{quote}

\vfill




\newpage



%
\pagestyle{plain}

\renewcommand{\theequation}{\thesection.\arabic{equation}}

\setcounter{equation}{0}

\section{Introduction}

The problem of coherent beam-beam interaction in storage ring colliders 
is one of the most important, and at the same time one of the most 
difficult problems in contemporary accelerator physics. Its importance 
lies in the fact that beam-beam interaction is the basic factor, 
limiting the luminosity of a circular collider. Nevertheless, some 
progress in the analytical treatment of the coherent beam-beam 
interaction has been made \cite{chao} - \cite{tzenov}, it is still far 
from being completely understood. In most of the references available 
the basic trend of analysis follows the perturbative solution of the 
Vlasov-Poisson equations, where the linearized system is cast in the 
form of an eigenvalue problem for the eigenmodes. 

An important question, which still remains unanswered is how to 
determine the invariant phase space density (equilibrium distribution 
function) if such exist. One possible way to approach this problem is 
to find an integral of motion (at least approximately) under certain 
conditions. Then the invariant density can be expressed as a generic 
function of the integral of motion. An attempt in this direction has 
been made by Alexahin \cite{alexahin}, who used the Deprit algorithm 
to determine the integral of motion (new action variable). 

In the present paper we develop a novel approach to the beam-beam 
interaction in circular colliders, based on the Renormalization Group 
(RG) method \cite{oono}. Originally this method has been proposed as 
a singular perturbation technique for differential equations. Naive 
perturbation expansions \cite{nayfeh} are well-known to produce 
secular terms, thus limiting the range of validity of the perturbation 
solution. The basic idea of the RG method is to remove secular or 
divergent terms by renormalizing the integration constants of the 
lowest order perturbative solution. Its extension to discrete 
symplectic maps is however not straightforward, and should be 
performed with care. Here we follow the regularization procedure 
outlined in the paper by Goto and Nozaki \cite{nozaki}. As shown in 
\cite{nozaki} the naive RG map, obtained as a result of 
renormalization of the lowest order solution preserves the symplectic 
symmetry only approximately, and does not describe the long-term 
behavior of the original map correctly. The symplecticity is recovered 
by a process of ``exponentiation'', yielding a symplectic RG map 
together with an explicit expression for the nonlinear tune shift. An 
alternative version of the RG method, based on the envelope technique 
\cite{kunihiro} has been applied to study non symplectic maps. 

The paper is organized as follows. In the next Section we derive the 
one-dimensional nonlinear beam-beam map. In Section 3 the regularized 
RG map and its integral of motion are obtained. The integral of 
motion thus found is further used in Section 4 to derive a set of 
coupled integral equations of Haissinski type for the invariant phase 
space density. 

\renewcommand{\theequation}{\thesection.\arabic{equation}}

\setcounter{equation}{0}

\section{The Nonlinear Beam-Beam Map}

We begin with the one-dimensional model of coherent beam-beam 
interaction in the vertical $(q)$ direction, described by the 
Hamiltonian 
\begin{eqnarray} 
H_k = {\frac {{\dot{\chi}}_k} {2}} {\left( 
p^2 + q^2 \right)} + \lambda_k 
\delta_p (\theta) V_k (q; \theta), 
\label{Hamiltonian} 
\end{eqnarray} 
\noindent 
where the normalized beam-beam potential $V_k (q; \theta)$ 
satisfies the Poisson equation 
\begin{eqnarray} 
{\frac {\partial^2 V_k} {\partial q^2}} = 
4 \pi \! \! \int \limits_{- \infty}^{\infty} 
\! \! dp f_{3-k} (q, p; \theta), 
\label{Poisson} 
\end{eqnarray} 
\noindent 
and 
\begin{eqnarray} 
\lambda_k = {\frac {R r_e N_{3-k} \beta_{kq}^{\ast}} 
{\gamma_{k0} L_{(3-k)x}}} 
{\frac {1 + \beta_{k0} \beta_{(3-k)0}} {\beta_{k0}^2}} 
\simeq {\frac {2 R r_e N_{3-k} \beta_{kq}^{\ast}} 
{\gamma_{k0} L_{(3-k)x}}}. 
\label{Lambda} 
\end{eqnarray} 
\noindent 
Here, $(k=1,2)$ labels the counter-propagating beams, $\theta$ is 
the azimuthal angle, ${\dot{\chi}}_k = R \beta_{kq}^{-1}$ is the 
derivative of the phase advance with respect to $\theta$, $R$ is 
the mean machine radius, $r_e$ is the classical electron radius, 
$N_{1,2}$ is the total number of particles in either beam, 
$\beta_{kq}^{\ast}$ is the vertical beta-function at the interaction 
point, and $L_{kx}$ is the horizontal dimension of the beam ribbon. 
In addition, the distribution function $f_k (q, p; \theta)$ is a 
solution to the Vlasov equation 
\begin{eqnarray} 
{\frac {\partial f_k} {\partial \theta}} + 
{\dot{\chi}}_k p {\frac {\partial f_k} {\partial q}} -  
{\frac {\partial H_k} {\partial q}} 
{\frac {\partial f_k} {\partial p}} = 0. 
\label{Vlasov} 
\end{eqnarray} 
\noindent 
In order to build the iterative beam-beam map we formally solve the 
Hamilton's equations of motion
\begin{eqnarray} 
{\dot{q}} = {\frac {dq} {d \theta}} = 
{\dot{\chi}}_k p, \qquad \qquad 
{\dot{p}} = {\frac {dp} {d \theta}} = - 
{\dot{\chi}}_k q - \lambda_k \delta_p (\theta) 
V'_k (q; \theta), 
\label{Hamilton} 
\end{eqnarray} 
\noindent 
where the prime implies differentiation with respect to the spatial 
variable $q$. By defining the state vector 
\begin{eqnarray} 
{\bf z} = {\left( 
\begin{array}{c} 
q \\ 
p \end{array} 
\right)}, 
\label{Statevec} 
\end{eqnarray} 
\noindent 
we can rewrite Eq. (\ref{Hamilton}) in a vector form 
\begin{eqnarray} 
\dot{{\bf z}} = {\widehat{\bf K}} (\theta) 
{\bf z} + {\bf F} {\left( {\bf z}; \theta \right)}, 
\label{Vechamilt} 
\end{eqnarray} 
\noindent 
where 
\begin{eqnarray} 
{\widehat{\bf K}} (\theta) = {\left( 
\begin{array}{clcr} 
0 & {\dot{\chi}}_k \\ 
-{\dot{\chi}}_k & 0 
\end{array} \right)}, \qquad \quad \qquad 
{\bf F} {\left( {\bf z}; \theta \right)} = 
{\left( 
\begin{array}{c} 
0 \\ 
- \lambda_k \delta_p (\theta) V'_k (q; \theta) 
\end{array} 
\right)}. 
\label{Matrix} 
\end{eqnarray} 
\noindent 
Performing a linear transformation defined as 
\begin{eqnarray} 
{\bf z} = {\widehat{\cal M}} (\theta) 
\vec{\xi}, 
\label{Linear} 
\end{eqnarray} 
\noindent 
where the matrix ${\widehat{\cal M}}$ is a solution of the linear 
equation with a supplementary initial condition
\begin{eqnarray} 
\dot{\widehat{\cal M}} = {\widehat{\bf K}} (\theta) 
{\widehat{\cal M}}, \qquad \quad \qquad 
{\widehat{\cal M}} (\theta_0) = {\widehat{\bf I}}, 
\label{Linequation} 
\end{eqnarray} 
\noindent 
we write the equation for the transformed state vector $\vec{\xi}$ 
as follows: 
\begin{eqnarray} 
\dot{\vec{\xi}} = {\widehat{\cal M}}^{-1} (\theta) 
{\bf F} {\left( {\bf z}; \theta \right)}, 
\qquad \quad \qquad 
{\vec{\xi}} (\theta_0) = {\bf z}_0. 
\label{Transpvec} 
\end{eqnarray} 
\noindent 
Equation (\ref{Transpvec}) can be solved directly to give 
\begin{eqnarray} 
{\bf z} (\theta) = {\widehat{\cal M}} (\theta) 
{\bf z}_0 + \int \limits_{\theta_0}^{\theta} \! \! 
d \tau {\widehat{\cal M}} (\theta) 
{\widehat{\cal M}}^{-1} (\tau) 
{\bf F} {\left( {\bf z} (\tau); \tau \right)}. 
\label{Solution} 
\end{eqnarray} 
\noindent 
It can be easily checked that the matrix of fundamental solutions 
${\widehat{\cal M}}$ to the unperturbed problem is of the form 
\begin{eqnarray} 
{\widehat{\cal M}} (\theta) = {\left( 
\begin{array}{clcr} 
\cos {\left[ \chi_k (\theta) - \chi_k (\theta_0)  \right]} &  
\sin {\left[ \chi_k (\theta) - \chi_k (\theta_0)  \right]} \\ 
- \sin {\left[ \chi_k (\theta) - \chi_k (\theta_0)  \right]} &  
\cos {\left[ \chi_k (\theta) - \chi_k (\theta_0)  \right]} 
\end{array} \right)}, 
\label{Fundamental} 
\end{eqnarray} 
\noindent 
so that 
\begin{eqnarray} 
{\bf z} (\theta) = {\widehat{\cal M}} (\theta) 
{\bf z}_0 - \lambda_k \int \limits_{\theta_0}^{\theta} \! \! 
d \tau \delta_p (\tau) 
V'_k {\left( q (\tau); \tau \right)} 
{\left( \begin{array}{c} 
\sin {\left[ \chi_k (\theta) - \chi_k (\tau)  \right]} \\ 
\cos {\left[ \chi_k (\theta) - \chi_k (\tau)  \right]} 
\end{array} 
\right)}. 
\label{Solut} 
\end{eqnarray} 
\noindent 
Applying the above expression (\ref{Solut}) in a small 
$\varepsilon$-interval $\theta \in (\theta_0 - \varepsilon, 
\theta_0 + \varepsilon)$ around the interaction point (located at 
$\theta_0$) and then taking the limit $\varepsilon \rightarrow 0$ we 
obtain the kick map 
\begin{eqnarray} 
q_{\rm kick} = q_0, \qquad \quad \qquad 
p_{\rm kick} = p_0 - \lambda_k V'_k 
{\left( q_0 \right)}. 
\label{Kickmap} 
\end{eqnarray} 
\noindent 
In order to obtain the rotation map in between successive kicks we 
apply once again expression (\ref{Solut}) in the interval 
$\theta \in (\theta_0 + \varepsilon, \theta_0 + 2 \pi - \varepsilon)$ 
\begin{eqnarray} 
q_{\rm rot} = q_{\rm kick} \cos 2 \pi \nu_k + 
p_{\rm kick} \sin 2 \pi \nu_k, \nonumber \\ 
p_{\rm rot} = - q_{\rm kick} \sin 2 \pi \nu_k + 
p_{\rm kick} \cos 2 \pi \nu_k. 
\label{Rotatmap} 
\end{eqnarray} 
\noindent 
Combining Eqs. (\ref{Kickmap}) and (\ref{Rotatmap}) we finally arrive 
at the one-turn beam-beam map 
\begin{eqnarray} 
q_{n+1} = q_n \cos 2 \pi \nu_k + 
{\left[ p_n - \lambda_k 
V'_k {\left( q_n \right)} \right]} 
\sin 2 \pi \nu_k, \nonumber \\ 
p_{n+1} = - q_n \sin 2 \pi \nu_k + 
{\left[ p_n - \lambda_k 
V'_k {\left( q_n \right)} \right]} 
\cos 2 \pi \nu_k. 
\label{Beambeamap} 
\end{eqnarray} 
It is important to note that the one-turn beam-beam map 
(\ref{Beambeamap}) is symplectic, since its Jacobian determinant is 
equal to unity 
\begin{eqnarray} 
\det {\frac {\partial {\left( q_{n+1}, p_{n+1} \right)}} 
{\partial {\left( q_n, p_n \right)}}} = 
\det {\left( \begin{array}{clcr}  
\cos 2 \pi \nu_k - \lambda_k V''_k \sin 2 \pi \nu_k &  
\sin 2 \pi \nu_k \\ 
- \sin 2 \pi \nu_k - \lambda_k V''_k \cos 2 \pi \nu_k & 
\cos 2 \pi \nu_k \end{array} \right)} \equiv 1. 
\label{Jacobian} 
\end{eqnarray} 

\renewcommand{\theequation}{\thesection.\arabic{equation}}

\setcounter{equation}{0}

\section{Renormalization Group Reduction of the Beam- \\ 
Beam Map}

The one-turn beam-beam map, derived in the previous Section can be 
further simplified by eliminating the canonical momentum variable 
$p$ from (\ref{Beambeamap}). Multiplying the first of Eqs. 
(\ref{Beambeamap}) by $\cos 2 \pi \nu_k$, multiplying the second one 
by $- \sin 2 \pi \nu_k$, and summing the two equations up we find 
\begin{eqnarray} 
q_{n+1} \cos \omega_k - 
p_{n+1} \sin \omega_k = q_n, 
\label{Elimination} 
\end{eqnarray} 
\noindent 
where 
\begin{eqnarray} 
\omega_k = 2 \pi \nu_k. 
\label{Omega} 
\end{eqnarray} 
\noindent 
Using Eq. (\ref{Elimination}) we obtain a second order difference 
equation 
\begin{eqnarray} 
{\widehat{\cal L}} q_n = q_{n+1} - 
2 q_n \cos \omega_k + q_{n-1} = 
- \epsilon \lambda_k 
V'_k {\left( q_n \right)} 
\sin \omega_k, 
\label{Map} 
\end{eqnarray} 
\noindent 
where $\epsilon$ is a formal small parameter (set to unity at the end 
of the calculations), taking into account the fact that the beam-beam 
kick is small and can be treated as perturbation. 

Next we consider an asymptotic solution of the map (\ref{Map}) for 
small $\epsilon$ by means of the RG method. The naive perturbation 
expansion 
\begin{eqnarray} 
q_n = q_n^{(0)} + \epsilon  q_n^{(1)} + 
\epsilon^2  q_n^{(2)} + \cdots 
\label{Expansion} 
\end{eqnarray} 
\noindent 
when substituted into Eq. (\ref{Map}) yields the perturbation 
equations order by order 
\begin{eqnarray} 
{\widehat{\cal L}} q_n^{(0)} = 0, 
\label{Perturb0} 
\end{eqnarray} 
\begin{eqnarray} 
{\widehat{\cal L}} q_n^{(1)} = 
- \lambda_k V'_k {\left(  q_n^{(0)} \right)} 
\sin \omega_k, 
\label{Perturb1} 
\end{eqnarray} 
\begin{eqnarray} 
{\widehat{\cal L}} q_n^{(2)} = 
- \lambda_k q_n^{(1)} 
V''_k {\left(  q_n^{(0)} \right)} 
\sin \omega_k, 
\label{Perturb2} 
\end{eqnarray} 
\begin{eqnarray} 
{\widehat{\cal L}} q_n^{(3)} = 
- \lambda_k {\left[ 
{\frac { q_n^{(1){\bf 2}}} {2}} 
V'''_k {\left(  q_n^{(0)} \right)} + 
q_n^{(2)} V''_k {\left(  q_n^{(0)} \right)} 
\right]} \sin \omega_k, 
\label{Perturb3} 
\end{eqnarray} 
\noindent 
Solving Eq. (\ref{Perturb0}) for the zeroth order contribution we 
obtain the obvious result 
\begin{eqnarray} 
q_n^{(0)} = A_k e^{i \omega_k n} + {\rm c.c.} = 
2 {\left| A_k \right|} 
\cos {\left( \omega_k n + \phi_k \right)}, 
\label{Solution0} 
\end{eqnarray} 
\begin{eqnarray} 
p_n^{(0)} = i A_k e^{i \omega_k n} + {\rm c.c.} = 
- 2 {\left| A_k \right|} 
\sin {\left( \omega_k n + \phi_k \right)}, 
\label{Solutionp0} 
\end{eqnarray} 
\noindent 
where $A_k$ is a complex integration constant, whose amplitude and 
phase are ${\left| A_k \right|}$ and $\phi_k$ respectively. 

Let us assume for the time being that the beam-beam potential 
$V_k (q)$ is a known function of the vertical displacement $q$. In 
what follows it will prove efficient to take into account the fact that 
the beam-beam potential $V_k (q)$ is an even function of the 
coordinate $q$. Odd multipole contributions to $V_k (q)$ will give rise 
to a shift in the closed orbit, and can be easily incorporated in the 
the calculations presented below. It is straightforward to check that 
the Fourier image of the beam-beam potential $V_k (\lambda)$, defined 
as 
\begin{eqnarray} 
V_k (q) = {\frac {1} {2 \pi}} \int 
\limits_{- \infty}^{\infty} \! \! d \lambda 
V_k (\lambda) e^{i \lambda q}, \qquad \qquad 
V_k (\lambda) = \int \limits_{- \infty}^{\infty} 
\! \! d q V_k (q) e^{- i \lambda q}, 
\label{Fourier} 
\end{eqnarray} 
\noindent 
retains the symmetry properties of $V_k (q)$, that is:
\begin{eqnarray} 
V_k (- \lambda) = V_k (\lambda). 
\label{Symmetry} 
\end{eqnarray} 
\noindent 
Using the expansion \cite{grad,stegun} 
\begin{eqnarray} 
e^{i z \cos \varphi} = \! 
\sum \limits_{m=- \infty}^{\infty} \! \! 
i^m {\cal J}_m (z) e^{i m \varphi}, 
\label{Expansio} 
\end{eqnarray} 
\noindent 
where ${\cal J}_m (z)$ is the Bessel function of the first kind of 
order $m$, and the explicit form of the zero order solution 
(\ref{Solution0}) we find 
\begin{eqnarray} 
V'_k {\left( q_n^{(0)} \right)} = 
\sum \limits_{M=1}^{\infty} {\cal C}_k^{(M)} 
A_k^{2M-1} e^{i (2M-1) \omega_k n} + {\rm c.c.} 
\label{Vkprime} 
\end{eqnarray} 
\noindent 
Here the coefficients ${\cal C}_k^{(M)}$ are functions of the 
amplitude ${\left| A_k \right|}$ and are given by the expression 
\begin{eqnarray} 
{\cal C}_k^{(M)} 
{\left( {\left| A_k \right|} \right)} = 
{\frac {1} {\pi}} {\frac {(-1)^M} 
{{\left| A_k \right|}^{2M-1}}} \int 
\limits_0^{\infty} \! \! d \lambda 
\lambda V_k (\lambda) {\cal J}_{2M-1} 
{\left( 2 \lambda {\left| A_k \right|} \right)}. 
\label{Coeffic} 
\end{eqnarray} 
\noindent 
Similarly for the second derivative of the beam-beam potential 
$V''_k {\left( q_n^{(0)} \right)}$, entering the second order 
perturbation equation (\ref{Perturb2}) we have 
\begin{eqnarray} 
V''_k {\left( q_n^{(0)} \right)} = 
{\cal D}_k^{(0)} + 
\sum \limits_{M=1}^{\infty} {\cal D}_k^{(M)} 
A_k^{2M} e^{i 2M \omega_k n} + {\rm c.c.}, 
\label{Vksecond} 
\end{eqnarray} 
\noindent 
where 
\begin{eqnarray} 
{\cal D}_k^{(0)} 
{\left( {\left| A_k \right|} \right)} = 
- {\frac {1} {\pi}} \int 
\limits_0^{\infty} \! \! d \lambda 
\lambda^2 V_k (\lambda) {\cal J}_0 
{\left( 2 \lambda {\left| A_k \right|} \right)}, 
\label{Coeffid0} 
\end{eqnarray} 
\begin{eqnarray} 
{\cal D}_k^{(M)} 
{\left( {\left| A_k \right|} \right)} = 
{\frac {1} {\pi}} {\frac {(-1)^{M+1}} 
{{\left| A_k \right|}^{2M}}} \int 
\limits_0^{\infty} \! \! d \lambda 
\lambda^2 V_k (\lambda) {\cal J}_{2M} 
{\left( 2 \lambda {\left| A_k \right|} \right)}. 
\label{Coeffid} 
\end{eqnarray} 
\noindent 
From the recursion property of Bessel functions \cite{grad,stegun} 
\begin{eqnarray} 
{\cal J}_{\nu - 1} (z) + {\cal J}_{\nu + 1} (z) = 
{\frac {2 \nu} {z}} {\cal J}_{\nu} (z) 
\label{Recursion} 
\end{eqnarray} 
\noindent 
we deduce an important relation to be used later 
\begin{eqnarray} 
{\cal D}_k^{(N)} - {\cal D}_k^{(N+1)} 
{\left| A_k \right|}^2 = {\left( 2N + 1 \right)} 
{\cal C}_k^{(N+1)}. 
\label{Important} 
\end{eqnarray} 
\noindent 
The solutions of the perturbation equations (\ref{Perturb1}) and 
(\ref{Perturb2}), taking into account (\ref{Important}) are given 
by 
\begin{eqnarray} 
q_n^{(1)} = {\frac {i \lambda_k n} {2}} 
{\cal C}_k^{(1)} A_k e^{i \omega_k n} + 
{\frac {\lambda_k \sin \omega_k} {2}} 
\sum \limits_{M=1}^{\infty} 
{\widetilde{\cal C}}_k^{(M+1)} A_k^{2M+1} 
e^{i (2M+1) \omega_k n} + {\rm c.c.}, 
\label{Solution1} 
\end{eqnarray} 
\begin{eqnarray} 
q_n^{(2)} = - {\frac {\lambda_k^2} {8}} 
{\cal C}_k^{(1){\bf 2}} 
{\left( n^2 + i n \cot \omega_k \right)} 
A_k e^{i \omega_k n} + 
{\frac {i \lambda_k^2 \sin \omega_k} {4}} n 
\sum \limits_{N=1}^{\infty} 
{\widetilde{\cal C}}_k^{(N+1)} 
{\cal D}_k^{(N)} {\left| A_k \right|}^{4N}
A_k e^{i \omega_k n} \nonumber 
\end{eqnarray} 
\begin{eqnarray} 
+ {\frac {i \lambda_k^2 \sin \omega_k} {4}} 
{\cal C}_k^{(1)} \sum \limits_{N=1}^{\infty} 
{\left( 2N + 1 \right)} 
{\widetilde{\cal C}}_k^{(N+1)} 
{\left[ n + i {\frac {\sin (2N+1)\omega_k} 
{\cos \omega_k - \cos (2N+1)\omega_k}} \right]} 
A_k^{2N+1} e^{i(2N+1) \omega_k n} \nonumber 
\end{eqnarray} 
\begin{eqnarray} 
+ {\frac {\lambda_k^2 \sin^2 \omega_k} {4}} 
{\cal D}_k^{(0)} \sum \limits_{N=1}^{\infty} 
{\frac {{\widetilde{\cal C}}_k^{(N+1)}} 
{\cos \omega_k - \cos (2N+1)\omega_k}} 
A_k^{2N+1} e^{i(2N+1) \omega_k n} \nonumber 
\end{eqnarray} 
\begin{eqnarray} 
+ {\frac {\lambda_k^2 \sin^2 \omega_k} {4}} 
\sum \limits_{M=1}^{\infty} 
\sum \limits_{N=1}^{\infty} 
{\frac {{\widetilde{\cal C}}_k^{(M+1)} 
{\cal D}_k^{(N)}} 
{\cos \omega_k - \cos [2(M+N)+1] \omega_k}} 
A_k^{2(M+N)+1} e^{i[2(M+N)+1] \omega_k n} 
\nonumber 
\end{eqnarray} 
\begin{eqnarray} 
+ {\frac {\lambda_k^2 \sin^2 \omega_k} {4}} 
{\sum \limits_{M=1}^{\infty}}^{\prime} 
{\sum \limits_{N=1}^{\infty}}^{\prime} 
{\frac {{\widetilde{\cal C}}_k^{(M+1)} 
{\cal D}_k^{(N)} {\left| A_k \right|}^{4N}} 
{\cos \omega_k - \cos [2(M-N)+1] \omega_k}} 
A_k^{2(M-N)+1} e^{i[2(M-N)+1] \omega_k n} + {\rm c.c.} 
\label{Solution2} 
\end{eqnarray} 
\noindent 
where the summation in the last term of Eq. (\ref{Solution2}) is 
performed for $M \neq N$, and 
\begin{eqnarray} 
{\widetilde{\cal C}}_k^{(N+1)} = 
{\frac {{\cal C}_k^{(N+1)}} 
{\cos \omega_k - \cos (2N+1) \omega_k}}. 
\label{Ctilde} 
\end{eqnarray} 
\noindent 
To remove secular terms, proportional to $n$ and $n^2$ we define the 
renormalization transformation $A_k \rightarrow {\widetilde{A}}_k (n)$ 
by collecting all terms proportional to the fundamental harmonic 
$e^{i \omega_k n}$ 
\begin{eqnarray} 
{\widetilde{A}}_k (n) = A_k + \epsilon 
{\frac {i \lambda_k n} {2}} 
{\cal C}_k^{(1)} A_k \nonumber 
\end{eqnarray} 
\begin{eqnarray} 
+ \epsilon^2 {\left[ - {\frac {\lambda_k^2} {8}} 
{\cal C}_k^{(1){\bf 2}} 
{\left( n^2 + i n \cot \omega_k \right)} + 
{\frac {i \lambda_k^2 \sin \omega_k} {4}} n 
\sum \limits_{N=1}^{\infty} 
{\widetilde{\cal C}}_k^{(N+1)} 
{\cal D}_k^{(N)} {\left| A_k \right|}^{4N} 
\right]} A_k. 
\label{Renampl} 
\end{eqnarray} 
\noindent 
Solving perturbatively Eq. (\ref{Renampl}) for $A_k$ in terms of 
${\widetilde{A}}_k (n)$ we obtain 
\begin{eqnarray} 
A_k = {\left[ 1 - \epsilon 
{\frac {i \lambda_k n} {2}} 
{\cal C}_k^{(1)} + 
O {\left( \epsilon^2 \right)} \right]} 
{\widetilde{A}}_k (n). 
\label{Unrenampl} 
\end{eqnarray} 
\noindent 
A discrete version of the RG equation can be defined by considering 
the difference 
\begin{eqnarray} 
{\widetilde{A}}_k (n+1) - 
{\widetilde{A}}_k (n)  = \epsilon 
{\frac {i \lambda_k} {2}} 
{\cal C}_k^{(1)} A_k \nonumber 
\end{eqnarray} 
\begin{eqnarray} 
+ \epsilon^2 {\left[ - {\frac {\lambda_k^2} {8}} 
{\cal C}_k^{(1){\bf 2}} 
{\left( 2n + 1 + i \cot \omega_k \right)} + 
{\frac {i \lambda_k^2 \sin \omega_k} {4}} 
\sum \limits_{N=1}^{\infty} 
{\widetilde{\cal C}}_k^{(N+1)} 
{\cal D}_k^{(N)} {\left| A_k \right|}^{4N} 
\right]} A_k. 
\label{Renequation} 
\end{eqnarray} 
\noindent 
Substituting the expression for $A_k$ in terms of 
${\widetilde{A}}_k (n)$ [see Eq. (\ref{Unrenampl})]  into the above 
Eq. (\ref{Renequation}) we can eliminate the secular terms up to 
$O{\left( \epsilon^2 \right)}$. The result is 
\begin{eqnarray} 
{\widetilde{A}}_k (n+1) = \left[ 1 + \epsilon 
{\frac {i \lambda_k} {2}} {\cal C}_k^{(1)} - 
\epsilon^2 {\frac {\lambda_k^2} {8}} 
{\cal C}_k^{(1){\bf 2}} 
{\left( 1 + i \cot \omega_k \right)} 
\right. \nonumber 
\end{eqnarray} 
\begin{eqnarray} 
\left. 
+ i \epsilon^2 
{\frac {\lambda_k^2 \sin \omega_k} {4}} 
\sum \limits_{N=1}^{\infty} 
{\widetilde{\cal C}}_k^{(N+1)} 
{\cal D}_k^{(N)} {\left| 
{\widetilde{A}}_k (n) \right|}^{4N} 
\right] {\widetilde{A}}_k (n). 
\label{Naivemap} 
\end{eqnarray} 
\noindent 
This naive RG map does not preserve the symplectic symmetry and does 
not have a {\it constant of motion}. To recover the symplectic 
symmetry we regularize the naive RG map by noting that the 
coefficient in the square brackets, multiplying ${\widetilde{A}}_k (n)$ 
can be exponentiated: 
\begin{eqnarray} 
{\widetilde{A}}_k (n+1) = 
{\widetilde{A}}_k (n) 
{\exp {\left[ i {\widetilde{\omega}}_k 
{\left( {\left| {\widetilde{A}}_k (n) 
\right|} \right)} \right]}}, 
\label{Symplmap} 
\end{eqnarray} 
\noindent 
where 
\begin{eqnarray} 
{\widetilde{\omega}}_k 
{\left( {\left| {\widetilde{A}}_k (n) 
\right|} \right)} = \epsilon 
{\frac {\lambda_k {\cal C}_k^{(1)}} {2}} + 
\epsilon^2 {\frac {\lambda_k^2} {8}}
{\left( - {\cal C}_k^{(1) {\bf 2}} 
\cot \omega_k +2 \sin \omega_k 
\sum \limits_{N=1}^{\infty} 
{\widetilde{\cal C}}_k^{(N+1)} 
{\cal D}_k^{(N)} {\left| 
{\widetilde{A}}_k (n) \right|}^{4N} 
\right)}. 
\label{Nonlintune} 
\end{eqnarray} 
\noindent 
It is clear now that the regularized RG map (\ref{Symplmap}) 
possesses the obvious integral of motion: 
\begin{eqnarray} 
{\left| {\widetilde{A}}_k (n+1) \right|} = 
{\left| {\widetilde{A}}_k (n) \right|} = 
{\sqrt{{\frac {J_k} {2}}}}. 
\label{Integral} 
\end{eqnarray} 
\noindent 
It is worthwhile to note that the secular coefficients of the 
$(2N+1)$-st harmonic $e^{i (2N+1) \omega_k n}$ can be summed up to 
give a renormalized coefficient, which expressed in terms of 
${\widetilde{A}}_k (n)$ does not contain secular terms. 

Proceeding in the same way as above, we can write the canonical 
conjugate momentum $p_n$ in the form 
\begin{eqnarray} 
p_n = i {\widetilde{B}}_k (n) 
e^{i \omega_k n} + {\rm c.c.} + 
{\rm higher \; harmonics}, 
\label{Momentum} 
\end{eqnarray} 
\noindent 
where 
\begin{eqnarray} 
{\widetilde{B}}_k (n+1) = 
{\widetilde{B}}_k (n) 
{\exp {\left[ i {\widetilde{\omega}}_k 
{\left( {\left| {\widetilde{A}}_k (n) 
\right|} \right)} \right]}}. 
\label{Symplmapb} 
\end{eqnarray} 
\noindent 
Using now the relation (\ref{Elimination}) between the canonical 
conjugate variables $(q, p)$ we can express the renormalized 
amplitude ${\widetilde{B}}_k (n)$ in terms of ${\widetilde{A}}_k (n)$ 
as 
\begin{eqnarray} 
{\widetilde{B}}_k (n) = i 
{\frac {e^{- i {\left( \omega_k + 
{\widetilde{\omega}}_k \right)}} - 
\cos \omega_k} {\sin \omega_k}} 
{\widetilde{A}}_k (n). 
\label{Relation} 
\end{eqnarray} 
\noindent 
Neglecting higher harmonics and iterating Eqs. (\ref{Symplmap}) and 
(\ref{Symplmapb}) we can write the renormalized solution of the 
beam-beam map (\ref{Beambeamap})
\begin{eqnarray} 
q_n = {\sqrt{2 J_k}} \cos \psi_k 
{\left( J_k; n \right)}, 
\label{Rensolutq} 
\end{eqnarray} 
\begin{eqnarray} 
p_n = \alpha_k {\left( J_k \right)} 
{\sqrt{2 J_k}} \cos \psi_k 
{\left( J_k; n \right)} - 
\beta_k {\left( J_k \right)} 
{\sqrt{2 J_k}} \sin \psi_k 
{\left( J_k; n \right)}, 
\label{Rensolutp} 
\end{eqnarray} 
\noindent 
where 
\begin{eqnarray} 
\psi_k {\left( J_k; n \right)} = 
{\left[ \omega_k + {\widetilde{\omega}}_k 
{\left( J_k \right)} \right]} n + 
{\widetilde{\phi}}_k, 
\label{Psi} 
\end{eqnarray} 
\begin{eqnarray} 
\alpha_k {\left( J_k \right)} = 
{\frac {\cos \omega_k - \cos {\left[ 
\omega_k + {\widetilde{\omega}}_k 
{\left( J_k \right)} \right]}} 
{\sin \omega_k}}, \qquad \qquad 
\beta_k {\left( J_k \right)} = 
{\frac {\sin {\left[ \omega_k + 
{\widetilde{\omega}}_k 
{\left( J_k \right)} \right]}} 
{\sin \omega_k}}. 
\label{Alphabeta} 
\end{eqnarray} 
\noindent 
It is easy to see that the integral of motion $J_k$ has the form of 
a generalized Courant-Snyder invariant and can be written as 
\begin{eqnarray} 
2 J_k = q^2 + {\frac {{\left[ p - 
\alpha_k {\left( J_k \right)} q \right]}^2} 
{\beta_k^2 {\left( J_k \right)}}}. 
\label{Invariant} 
\end{eqnarray} 
\noindent 
It is important to emphasize that Eq. (\ref{Invariant}) comprises a 
transcendental equation for the invariant $J_k$ as a function of the 
canonical variables $(q, p)$, since the coefficients $\alpha_k $ and 
$\beta_k$ depend on $J_k$ themselves. 

\renewcommand{\theequation}{\thesection.\arabic{equation}}

\setcounter{equation}{0}

\section{The Invariant Phase Space Density}

If an integral of motion $J_k$ of the beam-beam map (\ref{Beambeamap}) 
exists, it can be proved that the invariant phase space density 
$f_k^{(I)} (q, p)$ [which is a solution to the Vlasov equation 
(\ref{Vlasov})] is a generic function of $J_k$, that is 
\begin{eqnarray} 
f_k^{(I)} (q, p) = F_k {\left( J_k \right)} 
\qquad \quad \qquad 
(k = 1, 2). 
\label{Invdensity} 
\end{eqnarray} 
\noindent 
Here $F_k (z)$ is a generic function of its argument. Since the integral 
of motion $J_k$ is a functional of the invariant density of the opposing 
beam $f_{3-k}^{(I)} (q, p)$, Eq. (\ref{Invdensity}) comprises a coupled 
system of nonlinear integral equations for the invariant densities of the 
two counter-propagating beams. Let us find the integral of motion [see 
Eq. (\ref{Invariant})] up to first order in the perturbation parameter 
$\epsilon$. We have 
\begin{eqnarray} 
J_k = J_0 - {\frac {\lambda_k {\cal C}_k^{(1)} 
{\left( J_0 \right)}} {2}} 
{\left( p^2 \cot \omega_k + pq \right)}, 
\label{Invapprox} 
\end{eqnarray} 
\noindent 
where 
\begin{eqnarray} 
J_0 = {\frac {1} {2}} 
{\left( p^2 + q^2 \right)}. 
\label{Invar0} 
\end{eqnarray} 
\noindent 
The Fourier image of the beam-beam potential 
\begin{eqnarray} 
V_k (\lambda) = - {\frac {4 \pi} {\lambda^2}} 
\! \int \limits_{- \infty}^{\infty} \! \! dq' 
\! \! \int \limits_{- \infty}^{\infty} \! \! dp' 
f_{3-k}^{(I)} {\left( q', p' \right)} 
\cos \lambda q', 
\label{Image} 
\end{eqnarray} 
\noindent 
obtained by solving the Poisson equation (\ref{Poisson}) is next 
substituted into the corresponding expression [see Eq. 
(\ref{Coeffic})] for the coefficient 
${\cal C}_k^{(1)} {\left( J_0 \right)}$. Taking into account the 
recursion relation (\ref{Recursion}) as well as the identities 
\cite{grad} 
\begin{eqnarray} 
\int \limits_0^{\infty} \! dx 
{\cal J}_0 (x) \cos ax = 
{\frac {1} {\sqrt{1 - a^2}}} 
\qquad \quad \qquad 
[0 < a < 1], 
\label{Identity1} 
\end{eqnarray} 
\begin{eqnarray} 
\int \limits_0^{\infty} \! dx 
{\cal J}_2 (x) \cos ax = - 
{\frac {2 a^2 - 1} {\sqrt{1 - a^2}}} 
\qquad \quad \qquad 
[0 < a < 1] 
\label{Identity2} 
\end{eqnarray} 
\noindent 
we obtain 
\begin{eqnarray} 
{\cal C}_k^{(1)} {\left( J_0 \right)} = 
{\frac {8} {J_0}} \! \int 
\limits_{- \infty}^{\infty} \! \! dp' \! \! 
\int \limits_0^{\sqrt{2J_0}} \! \! dq' 
f_{3-k}^{(I)} {\left( q', p' \right)} 
{\sqrt{2J_0 - q'^2}}. 
\label{Coefficient1} 
\end{eqnarray} 
\noindent 
Thus, we finally arrive at the system of integral equations for the 
invariant phase space densities $f_k^{(I)} {\left( q, p \right)}$
\begin{eqnarray} 
f_1^{(I)} {\left( q, p \right)} = 
C_1 F_1 {\left[ J_0 - 
{\frac {4 \lambda_1} {J_0}} 
{\left( p^2 \cot \omega_1 + pq \right)} 
\! \int \limits_{- \infty}^{\infty} \! \! 
dp' \! \! \int \limits_0^{\sqrt{2J_0}} \! 
\! dq' f_2^{(I)} {\left( q', p' \right)} 
{\sqrt{2J_0 - q'^2}} \right]}, 
\label{Intequation1} 
\end{eqnarray} 
\begin{eqnarray} 
f_2^{(I)} {\left( q, p \right)} = 
C_2 F_2 {\left[ J_0 - 
{\frac {4 \lambda_2} {J_0}} 
{\left( p^2 \cot \omega_2 + pq \right)} 
\! \int \limits_{- \infty}^{\infty} \! \! 
dp' \! \! \int \limits_0^{\sqrt{2J_0}} \! 
\! dq' f_1^{(I)} {\left( q', p' \right)} 
{\sqrt{2J_0 - q'^2}} \right]}, 
\label{Intequation2} 
\end{eqnarray} 
\noindent 
where 
\begin{eqnarray} 
C_k = {\left[ \! \int \limits_{- \infty}^{\infty} 
\! \! dp \! \! \int \limits_{- \infty}^{\infty} 
\! \! dq F_k (q, p) \right]}^{-1}. 
\label{Normal} 
\end{eqnarray} 

It is instructive to calculate the first order nonlinear incoherent 
beam-beam tune shift. According to Eq. (\ref{Nonlintune}) we have 
\begin{eqnarray} 
{\widetilde{\omega}}_k^{(1)} 
{\left( J_k \right)} = 
{\frac {4 \lambda_k} {J_k}} \! \int 
\limits_{- \infty}^{\infty} \! \! dp' \! \! 
\int \limits_0^{\sqrt{2J_k}} \! \! dq' 
f_{3-k}^{(I)} {\left( q', p' \right)} 
{\sqrt{2J_k - q'^2}}. 
\label{Firstune} 
\end{eqnarray} 
\noindent 
Since we are interested in the first order 
$O{\left( \lambda_k \right)}$ contribution, we substitute in Eq. 
(\ref{Firstune}) the unperturbed phase space density 
\begin{eqnarray} 
f_k^{(I)} {\left( q, p \right)} = 
{\frac {1} {2 \pi \sigma_k^2}} \: 
{\exp \! {\left( - {\frac {p^2 + q^2} 
{2 \sigma_k^2}} \right)}}. 
\label{Unperturbed} 
\end{eqnarray} 
\noindent 
Simple manipulations yield 
\begin{eqnarray} 
{\widetilde{\omega}}_k^{(1)} 
{\left( J_k \right)} = 
{\frac {2 \lambda_k} 
{\sigma_{3-k} \sqrt{2 \pi}}} 
\exp {\left( - {\frac {J_k} 
{2 \sigma_{3-k}^2}} \right)} \! 
\int \limits_0^{\pi} \! \! d \psi 
(1 + \cos \psi ) \exp 
{\left( {\frac {J_k} 
{2 \sigma_{3-k}^2}} \cos \psi 
\right)}. 
\label{Manipulation} 
\end{eqnarray} 
\noindent 
Taking into account the integral representation of the modified 
Bessel function ${\cal I}_n (z)$ (see e.g. Ref. \cite{stegun}) 
\begin{eqnarray} 
{\cal I}_n (z) = {\frac {1} {\pi}} 
\int \limits_0^{\pi} d \tau 
\cos (n \tau) e^{z \cos \tau}, 
\label{Modified} 
\end{eqnarray} 
\noindent 
we obtain 
\begin{eqnarray} 
{\widetilde{\omega}}_k^{(1)} 
{\left( J_k \right)} = 
{\frac {\lambda_k \sqrt{2 \pi}} 
{\sigma_{3-k}}} 
\exp {\left( - {\frac {J_k} 
{2 \sigma_{3-k}^2}} \right)} 
{\left[ {\cal I}_0 {\left( {\frac {J_k} 
{2 \sigma_{3-k}^2}} \right)} + 
{\cal I}_1 {\left( {\frac {J_k} 
{2 \sigma_{3-k}^2}} \right)} 
\right]}. 
\label{Tuneshift} 
\end{eqnarray} 
\noindent 
A similar expression for the incoherent beam-beam tune shift was 
obtained in \cite{yokoya1}. 

\section{Concluding Remarks} 

As a result of the investigation performed we have obtained a 
regularized symplectic RG beam-beam map, that describes correctly the 
long-time asymptotic behavior of the original system. It has been 
shown that the regularized RG map possesses an integral of motion, 
which can be computed to any desired order. The invariant phase space 
density (stationary distribution function) has been constructed as 
a generic function of the integral of motion, and a coupled set of 
nonlinear integral equations for the distributions of the two 
colliding beams has been derived. Based on the explicit form of the 
regularized RG map, the incoherent beam-beam tune shift has been 
computed to first order in the beam-beam parameter. 

It is worthwhile to note that the method presented here is also 
applicable to study the four-dimensional symplectic beam-beam map, 
governing the dynamics of counter-propagating beams in the plane 
transverse to the particle orbit. 

\subsection*{Acknowledgments}

It is a pleasure to thank Yunhai Cai and Y. Alexahin for helpful 
discussions concerning the subject of the present paper. This 
research was supported by the U.S. Department of Energy.

 

\end{document}